# Zero-Emission Delivery for Logistics and Transportation: Challenges, Research Issues, and Opportunities

Janfizza Bukhari, j.bukhari@ubc.ca, Abhishek G Somanagoudar, abhigs@student.ubc.ca, Luyang Hou*, luyang.hou@hotmail.com, Omar Herrera, omar.herrera@ubc.ca, Walter Mérida, walter.merida@ubc.ca

MeridaLabs, The University of British Columbia, Vancouver, BC, Canada

## Abstract

Greenhouse gas, produced from various industries such as Power, Manufacturing, Transport, Chemical, or Agriculture, is the major source of global warming. While the transport industry is among the top three major contributors, accounting for 16.2% of global emissions. To counter this, many countries are responding actively to achieve net or absolute zero-emission goals by replacing fossil fuel with renewable energy sources. In response to this initiative, this chapter provides a systematic review of the use of zero-emission vehicles for a specific use case of package delivery. It first compares different green delivery systems that use unmanned aerial vehicles, electric vehicles, and fuel-cell trucks for certain weight categories. Specifically, a coordination of unmanned aerial vehicle and ground-based electric truck envisions a new paradigm of ground-based zero-emission vehicles where unmanned aerial vehicles can fly in the air beyond the visual line of sight empowered by future-generation wireless technologies. The integration of zero-emission vehicles for package delivery will encounter many challenges in analyzing, modelling, planning, and designing a green logistics system. This chapter investigates these challenges in the adoption of zero-emission vehicles with the existing research issues from a technical, environmental, economic, and political point of view. In addition, this study also sheds a new research perspective on artificial intelligence and integrated solutions for zero-emission deliveries.

## Introduction

Zero-emission refers to a non-negative carbon emission on the environment. The zero-emission vehicles (ZEVs) operate without causing any harmful emissions to the atmosphere. In fact, the transportation industry contributes to 16.2% of total global greenhouse gas (GHG) emissions with carbon dioxide ($CO_2$) as a key contributor accounting for 6.85 billion tons emissions in 2020 (McCormick, Climate Trace, 2020). The emissions by road-cargo transportation account for 4.76% of these gases (Ritchie, 2020). These numbers are a concern for every species on earth and environmental policymakers, as emitted gases cause global warming (Mann, 2022) which leads to unforeseen climatic changes and natural disasters (UNO, 2022). To address these issues, the United Nations Framework Convention on Climate Change (UNFCCC) has set some benchmarks to mitigate the global climate change crisis by 2030 (UNO, 2022). The Conference of Parties (COP) hosted by the UNFCCC in Pairs and recently in Glasgow, ratified 197 countries, among them 83 countries that have committed to net-zero target (Climate Watch Data, 2022). The term *net-zero* is defined as balancing the amount of GHGs produced to the amount removed from the atmosphere, aiming at a sustainable development spread across different industries. Net-zero is essential for logistics systems to transport goods and materials in an environment-friendly way.

Traditional transport vehicles use internal combustion engines (ICE) powered by gasoline or diesel fuel, which is non-renewable sources of energy extracted from fossil fuels. It indicates that the non-renewable energy sources will run out or will not be replenished in our lifetimes — or even in many, many lifetimes ('Non-renewable energy'). These fuels are burnt in ICEs to convert heat energy into mechanical motion, during this process it releases GHGs which cause air pollution. This increased human activity behind the wheel has led to an increasing $CO_2$ emission and triggered the cascading effect of rising global temperature, resulting in severe climate change calamities (heat waves, famines, floods, etc.). Fig.1 shows the time-lapse of 5 decades indicating the rising global temperature, which then leads to heatwaves, forest fires, melting of glaciers, etc. The projected rise in temperature for the near future is indicated in Fig.2, making the earth a giant ball of heat and further accelerate these events of natural disasters frequently.

This chapter describes the green transportation systems by introducing various trending technologies, research, and challenges in the transport and logistics sectors. The second section introduces the concept of zero-emission delivery systems, followed by categorizing electric vehicles and electric trucks. The third section discusses the challenges and research issues in implementation for package deliveries. The fourth section presents the future scope of green logistics systems and potential research opportunities. The last section summarizes this entire chapter for the use of zero-emission vehicles as a sustainable solution in the green logistics industry.

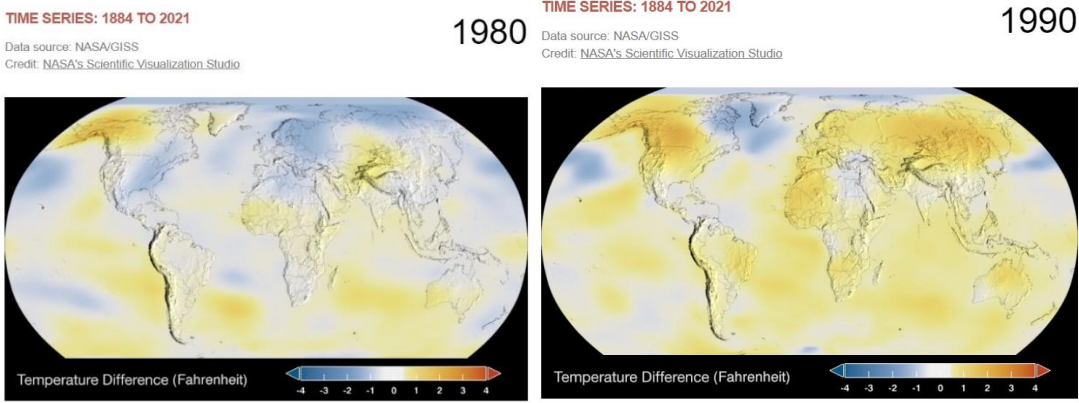

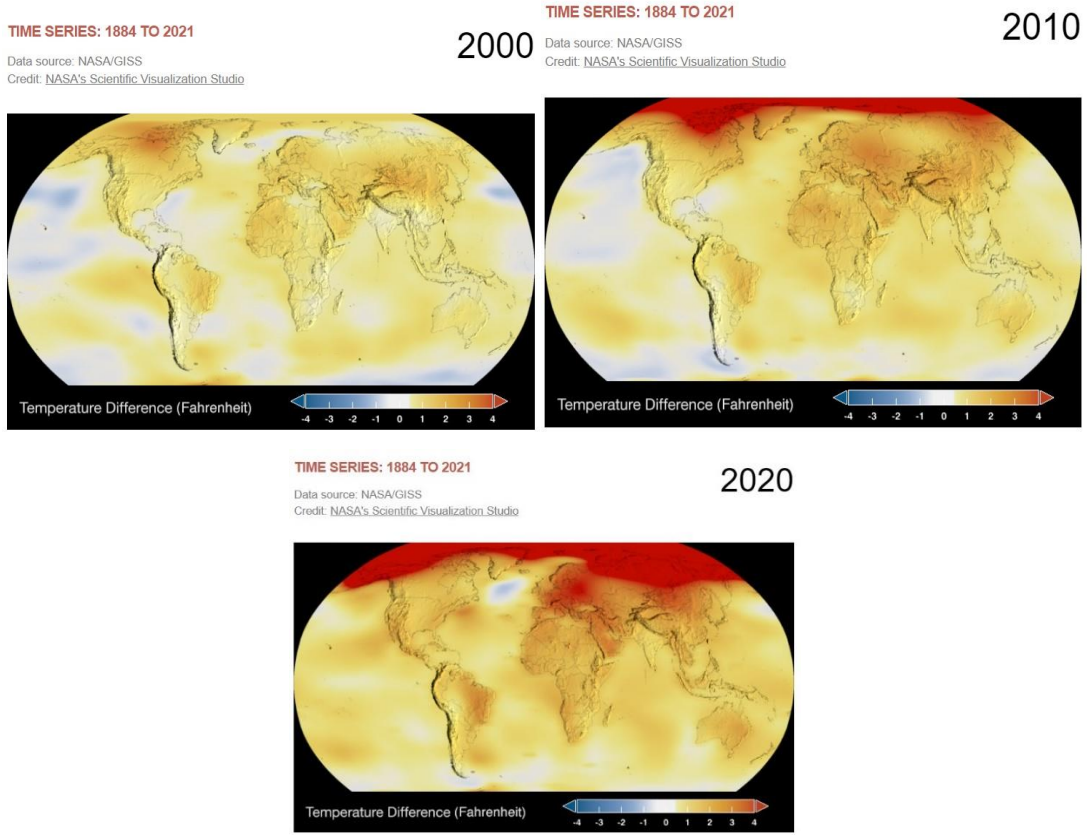

Fig 1 : Time series snippets from 1990-2020 for temperature difference (NASA, 2021).

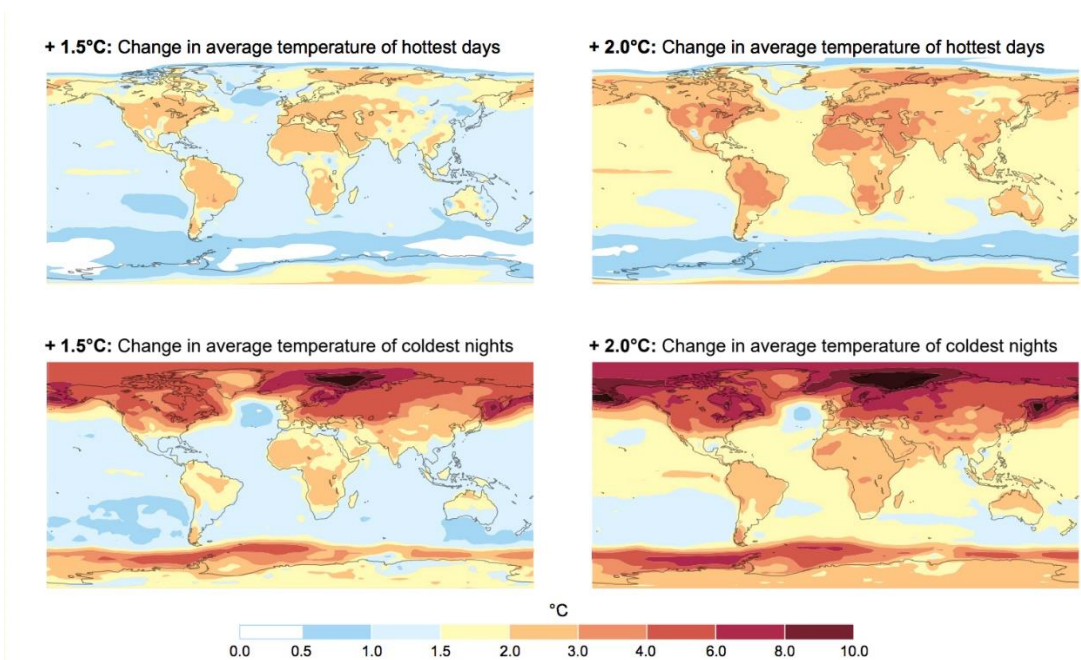

Fig. 2 : Projected temperature change effect in future if no measures are taken (NASA).

# Zero-Emission delivery options

Green logistics require replacing the conventional fossil fuel-driven vehicles with zero-emission vehicles in order to achieve environmental targets. To attain ecosystem sustainability, some of the alternatives to fossil fuel vehicles are electric vehicles (EVs) and other zero-emission vehicles (with hydrogen, solar, or wind energy as fuel source). In general, ZEVs can be classified into pure electric, battery powered, plug-in hybrid, hydrogen fuel cell powered vehicles. Each type has its own advantages and technical challenges, further detail of which is provided in later part of the chapter. These ZEVs not only include ground based heavy-duty vehicles but unmanned aerial vehicles (UAVs) like drones can also be considered as a mode of transportation for reducing $CO_2$ footprints in logistics sector. In line with global sustainability, the researchers are also aiming at developing a systems-based approach for integrating electric trucks (ET) and drones or UAVs for package delivery. The net-zero concept has a daunting prospect and to provide certainty about the pathway to enable full transition to zero-emission medium, mandatory measures in terms of investments and regulations are yet to be taken. This chapter aims at addressing net-zero emission in the logistics industry to counter and address a small percentage of change yet addressing the big picture when summed up. Several countries are enforcing the conversion of all road transport systems to 100% Electric Vehicles (EV) by 2040 (Fernández, Herrera and Mérida, 2020). Across the globe, research and integration of trending zero-emission technologies is being carried out which can be replicated in bigger cities (Merida Labs, 2020).

## Electric Vehicles

Electric vehicles can be classified into battery electric vehicle (BEV), plugin hybrid electric vehicle (PHEV) and hybrid electric vehicle (HEV) according to source of energy being used to power the vehicle. Table 1 provides a comparison of conventional and different ZEVs used for transportation purposes. The BEV runs on purely electric motors powered by batteries. PHEV are hybrid vehicles which operate on double power sources, an electric and a combustion engine. HEV is a fuel-driven machine with gasoline engine as a primary source of power. Both these types of hybrid vehicles use double power sources, however HEV cannot be plugged in to a power outlet for charging. Driving long distances requires more energy for which a greater number of batteries are needed, leading to increased vehicle weight. This increased battery and vehicle weight has variable effects on the efficiency and performance of the vehicle. These parameters have been addressed by different manufacturers and an optimal design of EVs are available for specific use. In comparison to ICEs, EVs are energy efficient (Kumar and Alok, 2020), but there are several challenges in the large-scale adaptation of EVs. Some of the factors influencing the large-scale adaptation include high battery cost, inadequate charging infrastructure, short-range (200 – 300 miles on a single charge) performance, and others. The EVs are further classified by type of vehicle in Table 1 as cars, bus, trucks, aircrafts, etc. The electric trucks have similar features for operations and performance varies with different weight categories and range of the vehicle. The recent advancement in automation, have added a new dimension to the safety, security, and reliability of these vehicles. The autonomy brings in great flexibility to drive these vehicles with or without human interventions providing efficiency in vehicle performance and increasing productivity in operations through a grid network.

*Table 1: Classification of transport vehicles by evolutions of engine or motor technology, purpose of vehicle utility, energy source used, type of fuel consumed, category or type of vehicles, and approximate tailpipe emissions during operation (Hong, Kuby and Murray, 2018; Khan et al., 2020; Litman, 2021; Pinto and Lagorio, 2022; Samarov and Verevkin, 2022).*

| Evolution of Engine/Motor Technology | Conventional | Hybrid | Plug-In Hybrid | Battery Electric | Other Zero Emission Vehicles |
|---|---|---|---|---|---|
| **Utility** | Personal/Public/Cargo Transport | Personal/Public Transport | Personal/Public | Personal/Public/Cargo Transport | Personal/Public/Cargo Transport |
| **Source of energy** | Petroleum Products (Fossil-Fuel) | Petroleum Products (Fossil-Fuel) | Petroleum Products (Fossil-Fuel) and Electricity | Electricity | Hydrogen/Electricity/Solar/CNG/Wind |
| **Consumption** | Gasoline/Diesel/LPG/Jet Fuel | Gasoline/Diesel/LPG/Jet Fuel + Battery | Gasoline/Diesel/LPG/Jet Fuel + Battery | Battery | Battery/Fuel Cells |
| **Type of Vehicles** | Car/Truck/Bus/Van/Aircraft/Ship/UAV/Others | Car/Truck/Bus/Van/Others | Car/Truck/Bus/Van/Others | Car/Truck/Bus/Van/Aircraft/UAV/Others | Car/Truck/Bus/Van/Aircraft/Ship/UAV/Others |
| **Approximate Tailpipe Emissions** | 100% | 50% | 25% | 0% | 0% |

## Unmanned Aerial Vehicles (UAVs)

UAVs are remotely or autonomously piloted mini aircraft systems. These are classified based on the weight and usage of the aircraft. Because of miniature and autonomy characteristics, UAVs are used in photography, videography, surveillance, geospatial, agricultural, and recently for transportation of goods (Moshref-Javadi and Winkenbach, 2021). These UAVs is powered by batteries, fuel cells, or fossil fuels. Most of the UAVs currently used are battery-operated. Similar to the autonomous ETs, UAVs also provide great flexibility in transportation of goods and services while avoiding busy road traffic. In the recent past, the UAV industry has seen tremendous growth for various civil applications. Among them delivery applications are a key focus to supplement present traffic and pollution scenarios. Several companies have patented technologies of drone-delivery designs and UAVs inside the warehouses (Moshref-Javadi and Winkenbach, 2021). These drones are battery-powered and have low carbon footprints compared to aircraft. Among them, Deutsche Post DHL group has experimented for more than four years on the last-mile package delivery. The term 'last-mile delivery' is meant for delivering the goods or packages to the customer or client at the doorstep from the warehouses. They used battery-powered UAVs and conventional transport trucks under the small-weight category of 0.5 to 25 kilograms of packages (Gary Mortimer, 2021). Currently, the last mile achieved by transport trucks globally, which are run on fossil fuels, thus causing higher $CO_2$ emissions, and hindering net-zero targets (Ritchie, 2020). The barriers in implementing these systems are the energy and cost-optimization model for last-mile delivery to maximize the profits for the cargo industries. Considering these challenges, this topic has gained popularity addressing ETs and UAVs (powered by Li-ion batteries and fuel-cell) to achieve sustainable cargo industry with zero-emission package delivery (Stolaroff *et al.*, 2018).

## Fuel-cell powered vehicles

Recent vehicles have implemented hydrogen as fuel and achieved a successful result in terms of operations, efficiency, and usability. This type of ZEVs has fuel-cell stack that uses hydrogen as an energy source to power the vehicle. Within the fuel-cell, chemical reaction takes place which produces heat and water, where water is disposed of through the tailpipe. Storing hydrogen within high pressure hydrogen tanks and its safety are some of the key challenges. In fact, fuel-cell does not store energy in it so it does not require recharging to generate power to drive the wheels but can be refueled in a short time. Compared to other ZEVs, hydrogen fuel-cell powered vehicles have lower GHG emissions; however, the hydrogen supply infrastructure cost is extremely high (Văn Ga *et al.*, 2022). The key contributing factor to this cost is the platinum required in the making of the fuel-cell system. Recent researches are focusing on hydrogen storage technologies including metal hydrides and storing hydrogen in liquid form which can significantly improve the energy density (Langmi *et al.*, 2002; Samarov and Verevkin, 2022).

Table 2 summarizes the advantages and disadvantages of using ground based electric vehicles and aerial vehicles for logistical operations. The lifecycle of zero-emission vehicles demands an efficient energy design based on emission free technologies where power is supplied to the vehicle before it gets discharged during logistical operation. Despite the recent trend of using environment-friendly electric vehicles for logistics industry, there is a need to understand the components and principles of technology to foresee the potential barriers on the way of their wide implementation.

*Table 2: Advantages and Disadvantages of UAVs and ground based electric vehicles in logistics operations* (Litman, 2021)

| Vehicles | Advantages | Disadvantages |
|---|---|---|
| UAVs | <ul><li>Compact and portable to operate</li><li>Autonomous</li><li>Zero tailpipe emission</li><li>Energy efficient</li><li>Programmed to operate simultaneously with multiple drones</li><li>No delays due to road traffic/blocks</li><li>Operational Cost efficient (specific use cases)</li></ul> | <ul><li>Limited range of operations (10-50km)</li><li>Limited weight caring capacity (0.5-25kg)</li><li>Dedicated Infrastructure required</li><li>Evolving Technology</li><li>Sophisticated Hardware, Software required and Battery Technology</li><li>High capital cost</li><li>Slow Return on Investment</li></ul> |

|  | • Low down time<br>• Last-Mile connectivity<br>• Less human intervention | • Restrictions for operations<br>• Complex operations and permissions approval process<br>• Vulnerable to Cyber attacks<br>• High battery charging time |
| --- | --- | --- |
| Ground based electric vehicles | • Zero tailpipe emission<br>• Energy efficient<br>• Possibility of being autonomous<br>• Cost efficient<br>• Environmentally sustainable<br>• Economically viable over long period | • High battery charging/down time<br>• Limited range of operations<br>• Robust Charging Infrastructure required<br>• Adaptation Challenges<br>• Sophisticated Hardware, Software required and Battery Technology<br>• High capital cost<br>• Slow Return on Investment<br>• Vulnerable to Cyberattacks |

## Challenges and Research Issues

At present, logistics sector uses ICE trucks, vans, cars, motorcycles, and bicycles to deliver parcel which require human intervention. The mode of delivery may depend upon their weight category and type of goods. There exist strong ties among the energy sources, public health expenditures, logistics operations and eco-environmental sustainability (Khan *et al.*, 2020). For a smooth adoption of ZEVs, key barriers include the upfront purchase cost, technology uncertainty, lack of charging infrastructure and public awareness and education. The logistics industry also witnessed either the lack of policies and programs or the implementation to help overcome a number of these barriers.

Innovations are posed with all-time micro-level challenges right from the ideation stage through experimentation, development, implementation, and adaptation. In addition, sustainable approach are always over arched by various other macro level factors like Environment, Economics, and Politics (Bours, Mcginn and Pringle, 2014). The below classifications help in addressing these issues in detail for implementation of green logistics. This section classifies all the three fundamentals and extends this characterization to discuss further on the challenges to adopt green logistics. The following will go from technical-environmental-economical-political dimension to analyze the challenges of adopting ZEVs in the transportation and logistics systems.

A. Technical

a. Operation and management

To power zero-emission ground vehicles, the energy conversion technologies rely heavily on special chemistry and materials necessary to facilitate the efficient charge transfer processes. So, the main challenge is the limited capacity of battery power of these electric vehicles which greatly restricts the execution of long-distance delivery tasks. HEVs are always on depleting mode which requires high power source; therefore, a special recharging mechanism is needed to automatically recharge the battery of them (Chellaswamy and Ramesh, 2017). Different operational models are presented by Moshref-Javadi et al. (Moshref-Javadi and Winkenbach, 2021) in order to provide exhaustive information and classification on drone delivery. These models include Pure-Drone (PD), Unsynchronized multi-modal (UM), Synchronized Multi-Modal (SM) and Resupply Multi-Modal (RM). The detailed analysis of these models has shown certain advantages such as easy deployment, extended service range, same day delivery, etc. However, there exists certain limitations as well for heavy and bulky parcels, including the congestion around depot, need of additional vehicles to carry parcels and sophisticated synchronization among them. The detailed reviewing of the papers has unfolded the vast research perspective on the ground based EVs, UAVs, their applications for package delivery addressing the advantages and limitations.

b. Green energy fuel

Energy production through renewable sources is termed as green energy. The main challenge with renewable fuel is the technology needed to produce the energy and availability of renewable resources to convert it into electric or heat energy. These two are the major challenges associated with it and once produced the storage, transportation, and distribution of these fuels like Natural gas, Hydrogen, Ammonia, etc. are other major factors for adapting these vehicles (Hydrogen Storage Challenges | Department of Energy; Langmi et al., 2002; Upchurch and Kuby, 2010).

c. Operating infrastructure

For seamless cargo delivery through zero-emission vehicles (both the ground and aerial vehicles) over a long-range, charging infrastructure plays a vital role. Currently, there are several academic research efforts addressing the charging infrastructure (recharging or refueling stations) based on mathematical models like Euclidean shortest path, mixed integer programming, heuristic approach, and spatial models (Upchurch and Kuby, 2010; Hong, Kuby and Murray, 2018; Dong *et al.*, 2019; Cokyasar, 2021a, 2021b; Pal, Bhattacharya and Chakraborty, 2021; Pinto and Lagorio, 2022) for electric trucks and drones. Liimatainen et al. (Liimatainen, van Vliet and Aplyn, 2019) envisions six different scenarios of electric truck and provides a comparison of the cargo transport data between two countries, their charging infrastructure, grid peak-load handling capabilities, type of goods handled and neglecting route optimization, truck-trailer configuration, type of trucks based on payload volume rather than weight classification. To achieve an energy-efficient delivery of goods, it suggests a medium sized truck with less than 26 tons of weight capacity to distribute parcels in Switzerland and Finland.

d. Large-scale production

The implementation and transitioning to green logistics are posed by the large-scale production and availability of these vehicles to meet the demand and compete with the cost per vehicle. In 2020, Europe saw a dip in adaptation of low-emission commercial vehicles due to high cost and low infrastructure issue associated with it (Zero-emission commercial vehicles in Europe | Statista, 2022). Thus, the large-scale production of these green transport

is must to suffice a market of 8.6 trillion U.S. dollars (Total global logistics market size by region 2020 | Statista, 2020).

B. Environmental

One of the main pillars of sustainability is the environment, issues addressed in this topic refer to reducing carbon footprints, and GHGs emissions for entire logistic systems. The macro level challenges in green logistics implementation are described below:

   a. Production and implementation

   The overarching challenge of adopting green logistics lie in the large-scale manufacturing as discussed in the above sections, but to setup and run this production facility is not simple, as it has to address all other issues like availability of the raw materials, feasibility of the production plant, labor cost, government policies, and mainly environmental policies set by respective nations. If all the above are addressed, then it circles back to is it environmentally sustainable, i.e., can this production and implementation setup run for generations to come without harming the environment by not causing any damage to biodiversity, ecosystem, climate, humans and other species on land, water, and air (Machrafi, 2012).

   b. Sustainability

   The production and implementation of green logistics should be sustainable i.e., if the production of such vehicles causes greater carbon footprint in other countries where it is manufactured and used in other country which cares about sustainability, then the global climate degradation stays the same and the term sustainability does not hold any meaning to it. Therefore, companies should consider a sustainable approach in manufacturing and implementation irrespective of region of production or implementation of these technologies (Abdalla *et al.*, 2018; Pfeifer, Prebeg and Duić, 2020; Sree Lakshmi, Rubanenko and Hunko, 2020). Some companies purchase carbon credits from other regions to offset their carbon footprints. This is one of potential solutions, but companies must be aggressive to drive such change.

C. Economical

   a. Production cost variables

The cost is key player in driving any economy, irrespective of technology if the cost of producing a green logistic vehicle is high then the logistic companies think twice to implement this technology until unless they are forced by external regulations such as policies and an environmental activist. The cost associated with setting up such infrastructure is also high which restricts the infrastructure providers from investing in such projects without guaranteeing a well-established ZEV market. Hence if there is a balance between policies and subsidies by the government then the adaptation transitions to smother phase despite variation in production costs.

   b. Implementation cost variables

As a last-mile delivery option, low altitude aerial vehicles can be deployed as a potential delivery option. The use of drones for package delivery with specific assumptions of capital and the operational cost is evaluated by Skoufi et al. (Skoufi, Evgenia, Filiopoulou, Evangelia, Skoufis, 2021). This work compares the drone package delivery with the motorcycle delivery system and provides a techno-economic analysis. The analysis proves that the operating cost in case of motorcycles is 1.5 times greater than that of drones, given a certain limitation on the drone usage. In addition, the initial investment cost for drones is higher compared to motorcycles, which are realized over a longer operational period. There is a need to standardize the charging and refueling infrastructure and its deployment cost.

   c. Government subsidies

The government policies and subsidies play a vital role in regulating and implementing new technologies like these and the cost associated with it cannot be borne by any individual logistics or private companies. The level of infrastructure needed to support ZEV use has not been properly addressed. From a consumer's perspective, public charging and refueling stations for these ZEVs are not yet deployed abundantly. Thus, there is a strong need for governments interventions and subsidies to scale these infrastructures.

   d. Switching and scrappage variables

Current vehicles need to be addressed with respect to the end of their life cycle, which causes greater environmental impacts from the scrappage of these vehicles. The study shows that considering environmental impact in scrappaging reduced the implementation of the process when compared to only stimulus package benefits (Li, Liu and Wei, 2013). Thus, these challenges lead to slower transitioning and adaptation of green logistics.

D. Political

UNO directed SDGs have put up sustainable goals, based on these countries have proactively drafted their own goals and pathways to implement net or absolute zero emissions.

   a. Region specific active policies, implementations

European nations have set their targets to 2030 to achieve net zero-emissions to keep the climate change within 2ºC. These targets include conversion of entire transportation industry to use renewable and zero emission fuel and technologies. On the other hand, without a reliable and widespread charging and refueling network of ZEVs, particularly in the case of countries where hydrogen fuel is scarce or inaccessible, consumers may not be willing to purchase a ZEV so both are inter-linked (Hardman et al., 2018).

   b. Region specific Passive Policies

Few nations abstain from the climate crisis scenario and thus these countries use a decade old technology in logistics and transportation which are obsolete and cause greater harm to global warming. These situations lead to greater challenges for other countries to cope with and to fight the global crisis collectively. Hence these are beyond the control of any researchers and UN bodies but can only be addressed by bigger nations and companies which can influence and drive the change. One such example is the world's largest trucking company DSV Panalpina, Denmark with a 47.7-billion-dollar market value (World's largest trucking companies by market value | Statista, 2021) and to address region wise then Asia-Pacific 3908.91 billion dollars in global logistic market (Total global logistics market size by region 2020 | Statista, 2020), such companies and region can drive change in the logistic market by adopting and promoting the green technologies.

However, most existing works focus on hypothetical models, with several restricted assumptions on technical parameters of vehicles (including the size, type e.g., ground vehicle or drone) and neglecting weather conditions, spatial data, social, environmental, economic, and political impacts for logistical operations. There is a need to standardize the charging and refueling infrastructure and its deployment cost. Furthermore, the existing literature related to package delivery utilizing clean energy vehicles identifies a few other challenges related to the operations and management, dispatching, scheduling, and routing of these ZEVs including drones.

The systematic analysis of some representative academic research is presented in Table 3. The results derived from evaluation of each paper have been segregated based on vehicle type, research area, recommendation of the researcher, and limitation or challenges mentioned above. The recommendation suggests few positive, negative, and limitations of the research which leads to further investigation of each case with predefined limitations of the research area. The research presented are in recent past due to the evolving technology and its adoption in the current market.

*Table 3: Classification of Research Articles by Vehicle Type, Research Area and Recommendation*

| Author | Vehicle Type | Research Area | Recommendation & Results | Limitations |
|---|---|---|---|---|
| (Liimatainen, van Vliet and Aplyn, 2019) | Electric Trucks | Study of potential use and conversion to ETs by 71% in Switzerland and 35% in Finland. Comparison between diesel and electric energy ratios. Study of market potential for types of ETs. | Medium duty <26t rigid ETs have high potential for implementation. Enables investment decisions on types of electric freight. | Region specific study (only 2 countries). |
| (Davis and Figliozzi, 2013) | Electric Trucks | Competitiveness by studying Routing, Speed, Energy, and cost model of diesel trucks and two specific models of ETs. | Key factors influencing includes routing, fleet size, battery life, range, payload, and purchase cost of ETs. | Battery life is assumed to be fixed over a lifespan. Resale of ETs deprecates to 20% after 10 years. Only considers 3 models of vehicles. |
| (Moshref-Javadi and Winkenbach, 2021) | Drone/UAVs | Study on drone-based logistics. Multi visit multi drones, Synchronized Multi-model with drone and a truck. | Models mainly recommended for e-commerce and healthcare applications. | Lack of academic study on food and mail delivery. |

| Reference | Vehicle Type | Description | Key Findings | Limitations |
|---|---|---|---|---|
| (Sawadsitang et al., 2019) | Drone/UAVs | Multi-objective and three stage stochastic optimization for drone delivery scheduling. | Trade-off between drone and another carrier service is recommended. | Limitation in flying conditions and distance travelled by drones. |
| (Figliozzi, 2020) | UAVs, Sidewalk Delivery Robots and Road Delivery Robots | Efficiency analysis of autonomous aerial and ground vehicles w.r.t vehicle-miles, energy consumption and CO2 emissions. | Autonomous delivery vehicles are efficient than ETs. All 3 types reduce carbon emissions. | Efficient and reduces emissions only for specific use case. No time reduction in on road travel for ground robots. New challenges with policies for labor market. |
| (Figliozzi, 2018) | Drone/UAVs | Includes economics of vehicle, battery, labor, and energy costs. | UAVs have low CO2 emissions per unit distance. | Speed and reliability. Payload and limited range. Relatively high vehicle phase emissions. |
| (Skoufi, Evgenia, Filiopoulou, Evangelia, Skoufis, 2021) | Drone/UAVs | Techno economic analysis of package delivery by drone. Involves cost estimates for specific drones. | Results show motorcycle operating costs are 1.5 times higher than drone costs. | Study restricted to specific region and pertained to only cost and technical factors. Limitation in comparison with motorcycle. |
| (Kirschstein, 2020) | Drone/UAVs & Electric Vehicles | Energy demand of drones for deliveries based on environmental conditions and flight patterns. Comparison with Diesel and Electric Truck. | Sole drone-based delivery is not recommended with energy perspective in Urban area. | Set customer and path for package delivery. Technical more energy required except for moderate environmental conditions. |
| (Eftekhari, 2019) | Electric Vehicles and Lithium-ion Batteries (LiBs) | Addresses availability of battery raw material(lithium). Study to go beyond LiBs. | Transition to EVs is a must. LiBs are safest for EVs. Develop new Battery systems. | Limitations in research strategies. Policies and resource shortage of raw materials. Economic factors on research of EVs. |
| (Fernández, Herrera and Mérida, 2020) | Electric Vehicles | Machine learning based EV charging stations demand. Energy demand model to estimate and balance load during peak and off-peak hours. | Peak load can fluctuate from 1 to 3 times during high-consumptions weekdays. | Region specific study with limited charging stations. Infrastructure and technical challenges. |

| Reference | Topic | Description | Results | Notes |
|---|---|---|---|---|
| (Hong, Kuby and Murray, 2018) | Drone Charging Stations | Recharging station coverage model for drone delivery service planning *Euclidean Shortest Path (ESP) *Mixed Integer Programming (MIP) *Heuristic Solution. | With the flight range doubled, 7 stations can cover 97.6% of total demand, while the same level of coverage requires more than 30 stations with the shorter-range parameter. | *Technological Factors *Weather Conditions * Payload * Recharged as many times as needed to reach their destination (one station to other until destination) * Return to Warehouse/stations after delivery without running out of fuel *Assumption based on Amazon Manager Saying 10 miles drone coverage diameter for delivery. |
| (Pinto and Lagorio, 2022) | Drone Charging Stations | The number of charging stations to be deployed, minimizing the travelled distances •Aims at defining a network of charging stations connecting customers delivery points to hubs minimizing the service costs (i.e., the total distance travelled by drones), the infrastructural investments (i.e., the number of charging points) or a combination of the two using a bi-objective optimization model. | This result supports the necessity of a fast and reliable heuristic method. *The HEURISTIC method is less sensitive to the network size. This result would support the use of the HEURISTIC for larger instances to test different alternatives in a relatively short time *For example, parallelizing the paths generation can drastically reduce the required times. In contrast, the minimization of the number of active stations with the OPTIMAL model (case with θ = 0.0) usually requires significant amount of time (hours rather than seconds). | Rapidly charging/changing the batteries is available *Offer enough space to accommodate a sufficiently large number of drones simultaneously *Objective: the minimization of the drones operating cost, under the assumption that (a portion of) the cost depends upon the travelled distances. The second objective is the minimization of the number of charging stations installed, under the constraint that all delivery points are covered and connected via a path to a hub. |

In a green logistic ecosystem, control layer is also important comprising of a communication and control setup to coordinate these ZEVs including flying drones. This synchronization is a must for reliable routing and scheduling of logistical tasks. The information and communication technologies can be exploited in transportation sector to move towards the low-carbon economy preventing further climate change (Decarbonizing industries with connectivity & 5G - Ericsson, 2021). Using next-generation wireless technologies like 5G and beyond within the network composed of electricity distribution grids and ZEVs, digital transformation efforts can be accelerated with improved operational efficiencies because of their capability to provide global connectivity with great speed and lower latency with affordable costs. Besides, for a better estimation of economic benefits, variables like expenditure on energy and the external health costs cannot be ignored while assessing the worth of different zero-emission vehicles. The higher cost of ZEVs and lack of equivalent models (e.g., pickup trucks) to meet the required performance, lack of access to maintenance and repair are few other prospective challenges, green logistics industry could face these days.

# Opportunities

To combat climate changes and environmental pollution, sustainable transport particularly electrification of the transport sector is essential. However, several barriers including social, economic, and technological challenges still exist, and researchers are trying to explore the possible opportunities to resolve these issues. The synergetic development of ZEVs and the green logistics is raising many promising practices.

Zero-emission vehicles are operated on energy generated by fuels without GHG emission, and these fuels can be extracted from non-renewable or renewable source. The potentials of renewable energy sources including hydro power, wind power, solar power, and bio power can also be utilized to satisfy growing energy demands of logistics sector. Among several other renewable energy resources, solar is identified as the main source for power generation. Work (Pfeifer, Prebeg and Duić, 2020) investigates the challenges and opportunities for development of zero emission ferry lines between islands where energy systems are connected to the electricity grid. The same concept can be adopted for ground-based ZEVs used in green logistics. From the economic perspective, it is feasible to connect solar power plant or wind turbine power plant to the delivery and transportation system, ensuring cleanest power for charging of ZEVs. However, implementing several renewable energy sources in logistics involve certain risk factors including high financial investments (Giera and Kulińska, 2021). Proper energy generation for charging purposes is essential and location of such sources also matters for the efficiency of ZEVs (Sree Lakshmi, Rubanenko and Hunko, 2020). Integrating renewable energy with ZEVs is a potential alternative to scarce non-renewable sources, which will play its role in improving the growth of the ZEV market.

Other than the environmental benefits, ZEVs can achieve better performances for package delivery since the highly efficient electric motors provide high torque with low rotations per minute. For maximizing the profits even further in performing green logistical operations, a cooperative strategy can also be adopted by using integrated zero-emission truck-drone based delivery system as a last-mile option (Jackson and Srinivas, 2021). The effectiveness of such an integrated delivery system can be justified in terms of reduced overall delivery time and energy (Ferrandez *et al.*, 2016). There could be other plausible scenarios of using unmanned aerial vehicles like drones with a zero-emission truck to distribute the parcels to different locations (Ham, 2018). One possibility could be utilizing multiple drones per truck for delivery in remote areas for beyond visual line of sight logistical operations. Determining the best route in case of single drone per truck or even in case of multiple zero-emission trucks and multiple drones, efficient route planning is another area of research (Gonzalez-R *et al.*, 2020). In such cooperative and green delivery system, scheduling of drone missions beyond visual line of sight is also challenging other than routing (Amorosi *et al.*, 2020; Baek *et al.*, 2020) and there should be a cost-effective approach based on the ratio between the truck and drone speed and time needed by launching and recovery (Crişan and Nechita, 2019).

The urban transportation and logistics sector is under intensifying pressure in delivering a better service at an ever-lower costs. For green logistics, both the frequency and the duration of charging or refueling are concerning. Without careful operating and management, on-route charging for logistics vehicles may increase fuel costs and reduce the economic attractiveness (He, Liu and Song, 2020). Therefore, the future logistics system will integrate ZEVs, logistics companies, charging or refueling facilities, and customers in a more collaborative way. Moreover, the implementation of green logistics provides more opportunities for ZEVs to be incorporated with government policy goals, land use planning, urban design, and associated system management as integral components of the overall infrastructure design.

Artificial intelligence (AI) is already at the forefront of driving valuable strategies optimizing the green logistics systems (Poole and Mackworth, 2010). The advent and impacts of AI technologies and big data is evidencing the benefits of its implementation in facilitating missions for achieving smart, sustainable logistics and net-zero commitments. Also, the massive data exchanging between different subsystems in green logistics requires extensive communication efforts, thus making different sectors execute at high frequencies. Studying and understanding how to build, analyze, model, and manage a green logistics system is a strong future research direction, as well as developing integrated solutions that are capable of dealing with the aforementioned challenges leading to efficiency, economics, and sustainability, which benefits not just the academia but also the wider society served by the transportation and logistics system. Moreover, green logistics allows ZEVs to interact with the power system more proactively through, for instance, V2X paradigm, i.e., vehicle to grid (V2G), vehicle to vehicle (V2V), vehicle to building (V2B), and so on (IEA, 2020).

The government should also play its role in policy making to facilitate the ZEV penetration in the logistics systems. For promoting the adoption of renewable energy sources for logistical operations, loans should be offered to the corporate sector, which can in turn increase foreign direct investment inflows. Such measures are able to be taken by close collaboration between the regulatory authorities and the logistics industry. The government should offer subsidies on ZEVs and reduce tax on green vehicles, which will help to reduce the carbon emissions into the atmosphere, promoting the sustainable agenda.

## Summary

This chapter provides an overall perspective of deploying ground-based electric vehicles, UAVs, and other zero-emission transport vehicles for package delivery to desired customers. The intention is to establish a clean and green logistic system offering zero-emission delivery of goods satisfying the sustainable development goals. The research efforts indicate the adoption of medium sized ground electric vehicles and UAVs with limited weight carrying capacity and travel distance would serve the green package delivery system efficiently compared to ICEs. The current work is a motivation to address the challenges in deploying the eco-friendly transport vehicles for sustainable package delivery. The future scope can be extended to achieve an optimal green solution to build a connected network of multi-mode delivery systems. This objective of optimization can be achieved by defining the size of a zero-emission vehicle (ground vehicle or drone), type of batteries or fuel cells (for drones and trucks), alert system on charging or refueling schedule, and placement of these refuel stations (Figliozzi, 2018, 2020; Liimatainen, van Vliet and Aplyn, 2019; Bonsu, 2020; Transport Canada, 2021). These are a few important and technically challenging parameters to be addressed for effective using these integrated delivery systems in an eco-friendly environment.

# List of Abbreviations

| **BEV** | Battery Electric Vehicle |
|---|---|
| **COP** | Conference of Parties |
| **$CO_2$** | Carbon dioxide |
| **DHL** | Deutsche Post |
| **EV** | Electric Vehicle |
| **ET** | Electric Trucks |
| **HEV** | Hybrid Electric Vehicle |
| **PD** | Pure Drone |
| **PHEV** | Plug-in Hybrid Electric Vehicle |
| **R&D** | Research and Development |
| **RM** | Resupply Multi-Modal |
| **SM** | Synchronized Multi-Modal |
| **UAV** | Unmanned Aerial Vehicle |
| **UNO** | United Nations Organization |
| **UNFCCC** | United Nations Framework Convention on Climate Change |
| **UM** | Unsynchronized multi-modal |
| **UPS** | United Parcel Service |
| **ZET** | Zero emission trucks |
| **ZEV** | Zero emission vehicles |